\input nature_pp.sty
\input psfig

\baselineskip=20.0pt

\def\gsim{\;\rlap{\lower 2.5pt
 \hbox{$\sim$}}\raise 1.5pt\hbox{$>$}\;}
\def\lsim{\;\rlap{\lower 2.5pt
   \hbox{$\sim$}}\raise 1.5pt\hbox{$<$}\;}

\centerline{\bfb Viscosity Mechanisms in Accretion Disks}

\medskip
\centerline{\bfb Kristen~Menou$^{1,2}$}
\smallskip    

\bigskip

{\ssbig The self-sustained turbulence which develops in magnetized
accretion disks is suppressed in the weakly-ionized, quiescent disks
of close binary stars. Because accretion still proceeds during
quiescence, another viscosity mechanism operates in these systems. An
anticorrelation of the recurrence times of SU UMa dwarf novae with
their mass ratio supports spiral waves or shock-waves tidally induced
by the companion star as the main process responsible for accretion in
the quiescent disks.  Other weakly-ionized gaseous disks in systems
lacking a massive companion have to rely on yet another transport
mechanism or they could be essentially passive.  }
\medskip
\footnote{}{$^1$Princeton University, Department of Astrophysical
Sciences, Princeton N.J. 08544, USA.}
\footnote{}{$^2$Chandra Fellow}

Accretion discs{\it(\Ref{pri81})} are present in a variety of
astrophysical objects, such as mass-transfer binaries, young stellar
systems and active galactic nuclei{\it(\Ref{fkr92})}. The main
uncertainty on the structure of accretion disks is the nature and
magnitude of their viscosity{\it(\Ref{ss73})}, the process by which
the gas loses angular momentum to accrete onto the massive central
object. Theoretical studies of magnetized accretion disks have shown
that magneto-hydrodynamical (MHD) turbulence provides the necessary
outward angular momentum transport for accretion to
proceed{\it(\Ref{bh91},\Ref{bh99})}. The problem of identifying a
viscosity mechanism can also be addressed empirically, by constraining
the magnitude of viscosity in unsteady disks, where it directly
relates to the observed variability{\it(\Ref{fkr92})}. This is best
done in transient close binaries, such as recurrent dwarf novae, with
geometries, masses and mass accretion rates that are relatively well
known{\it(\Ref{war95})}.

The transient nature of accretion in the disks of close binaries is
understood as a thermal-viscous limit cycle due to sudden changes of
the disk opacity when hydrogen recombines{\it(\Ref{meys81})}. Global
disk evolution models reproduce the main properties of observed
outbursts, but only if the efficiency of transport and dissipation is
less in quiescence than during
outburst{\it(\Ref{sma84},\Ref{can93})}. This requirement is consistent
with the low level of ionization predicted in quiescence. In terms of
the magnetic Reynolds number
$$
Re_{\rm M} \equiv C_s H / \eta,
$$
where $C_s$ is the sound speed, H the disk scale height and $\eta$ the
resistivity (inversely proportional to the ionization fraction), local
shearing-box MHD simulations{\it(\Ref{hgb96},\Ref{flem00})} predict a
reduced efficiency of angular momentum transport when $Re_{\rm M}
\lsim 10^4$, a value comparable to what is expected in quiescent
disks{\it(\Ref{gm98})}. It is unclear, however, whether MHD turbulence
is still responsible for transport during this phase, at a reduced
level, or is entirely suppressed.

This question is addressed here by using results from shearing-box
simulations as an input for a global numerical model designed to study
disk instabilities{\it(\Ref{ham98})}. The model provides reliable
values for the temperature, density, and therefore $Re_{\rm M}$,
throughout the disk (thermal ionization and resistive diffusion are
the dominant processes in the disks considered
here{\it(\Ref{gm98})}). The viscosity parameter $\alpha$ used in the
model (a measure of the disk viscosity in units of $C_s H$) is
tabulated as a function of $Re_{\rm M}$ according to the most recent
(zero net flux) MHD simulations that include resistive
effects{\it(\Ref{flem00})}. The decay of $\alpha$ is locally limited
to an e-folding time-scale corresponding to $4$ orbits, to allow for
the finite time of field resistive diffusion. Other model parameters
are chosen to represent the disk around the white dwarf of the
prototypical dwarf nova SS Cyg{\it(\Ref{gm98})}: mass of the primary
$M_1=1.2~M_\odot$, disk inner and outer radii $R_{\rm in} = 5 \times
10^8$~cm and $R_{\rm out} = 4 \times 10^{10}$~cm, and mass transfer
rate of the companion $\dot M_T= 10^{-9}~M_\odot$~yr$^{-1}$.

The evolution of a disk annulus as it goes into quiescence in a
standard model is represented by triangles in Fig.~1. The evolution
when physically-motivated values of $\alpha$ are used (circles) shows
that the disk does not saturate at a low level of MHD turbulence, but
suffers a runaway cooling as dissipation and $Re_{\rm M}$ become less
and less important. Although the evolution was limited to a minimum
value $\alpha=10^{-4}$ and was stopped when $Re_{\rm M}$ reached
$100$, the disk was still cooling down at that point. The outcome is
independent of the value adopted for the field diffusion e-folding
time-scale, as long as it is short compared to the viscous time-scales
in the disk.  Although the presence of a secondary viscosity mechanism
could be responsible for a residual level of MHD turbulence, by
maintaining the disk at a temperature ($\sim 2000$~K) sufficient for
the field to remain coupled to the gas, the disappearance of
self-sustained MHD turbulence is guaranteed.

This is a conservative result, because not using a minimum value of
$10^{-4}$ for $\alpha$ would have lead to stronger turbulence
decay. Moreover, the enhanced reduction of transport and dissipation
shown by higher resolution MHD
simulations{\it(\Ref{hgb96},\Ref{flem00})} further supports the case
for turbulence decay.  The runaway decay is found at other radii in
the disk as well (the value of $Re_{\rm M}$ tends to be smaller at
smaller radii because of the nearly flat profile of $T_{C}$ in
quiescence and the scaling of $Re_{\rm M}$ with $H$), and additional
models show that the same result holds for the disks of x-ray
transients{\it(\Ref{xrb95},\Ref{men00})}, around neutron stars and
stellar-mass black holes.

Accretion is known to occur during quiescence in transient close
binaries{\it(\Ref{war95})}, so that another transport mechanism
operates in the disks. The rapid disk expansion observed during the
outbursts of several eclipsing dwarf novae{\it(\Ref{war95})} is
consistent with MHD-driven accretion because it shows that disk
internal stresses dominate transport during this phase. On the other
hand, the same disks are observed to shrink between consecutive
outbursts, which is a signature that transport is dominated by the
tidal torque due to the companion star at least in the outer regions
of the disk during quiescence.

The theory of tidally induced spiral waves or shock-waves predicts a
reduced effective torque for smaller ratios of the companion mass to
the accretor mass{\it(\Ref{pp77},\Ref{gt80})}. Because the recurrence
times of dwarf novae represent the time-scales for mass and angular
momentum redistribution in the quiescent disks, an anticorrelation of
the recurrence times with mass ratios is expected if tidal torques
dominate transport in the quiescent disks{\it(\Ref{ls91})}. An
anticorrelation exists for a sample of SU UMa dwarf
novae\footnote{$^3$}{SU UMa type dwarf novae show characteristic, long
superoutbursts in addition to the more frequent normal outbursts of U
Gem type dwarf novae.}  (Fig.~2). This sample represents all SU UMa
stars with well known masses and recurrence
times{\it(\Ref{rkcat99})}. The anticorrelation, which is significant
for normal and super outbursts, supports tidal effects due to the
mass-transferring star as the dominant transport mechanism during
quiescence in transient close binaries.  In the same six systems, no
apparent correlation of $t_{\rm rec}$ or $T_{\rm rec}$ with the
orbital period is found. For U Gem type dwarf novae, another subclass
for which there are enough systems with well known masses and
recurrence times to repeat the exercise, no correlation of the
recurrence times with mass ratio or orbital period is found.

A possible reason why the relation is apparent only for SU UMa stars
is that an anticorrelation of recurrence times with the mass ratio $q$
can easily be masked by differences in the rates at which mass is
transferred by the companion, because these also influence the
frequency of outbursts. SU UMa stars may constitute an ideal subsample
of dwarf novae to test for such a relation because they mostly lie in
a small range of orbital periods, below the period
gap{\it(\Ref{war95})}. Mass transfer is thought to be driven by
gravitational radiation in these systems{\it(\Ref{king88})}, which
means that it should primarily depend on the geometry and masses of
the system.  The case of mass transfer driven by magnetic braking
(when the binary orbital angular momentum is lost via the wind of the
mass-transferring companion), which applies to most U Gem type dwarf
novae above the period gap, may be more complex. The much wider range
of orbital periods covered by U Gem stars may also contribute to the
absence of a relation for these systems.

Differences in the mass transfer rates of SU UMa stars themselves
could explain the relation (Fig.~2). However, in the limit $q << 1$
and under the assumption that the secondary star fills its Roche lobe,
predictions for the mass transfer rate due to gravitational radiation
give $\dot M_T \propto q^2 M_1^{8/3}$, where $M_1$ is the accretor
mass{\it(\Ref{kkl96})}. There is no evidence for a dependence of the
recurrence times with $M_1$ in the data set considered here, as
expected if the primary reason for the anticorrelation is a reduced
efficiency of transport for smaller values of $q$.

The anticorrelation is consistent with a more general, qualitative
trend among transient close binaries for longer recurrence times in
systems with smaller mass ratios (omitting possible complications due
to evolutionary issues and differences in the mass transfer rates or
the accretion flow structures). This is the case of WZ Sge type dwarf
novae, which have longer recurrence times and lower mass ratios than
the rest of the dwarf nova population{\it(\Ref{war95})}.  Similarly,
soft x-ray transients containing accreting black holes, as a class,
have longer recurrence times than similar systems containing less
massive, accreting neutron stars{\it(\Ref{xrb95})}.

Theoretically, it has been argued that the tidal perturbations due to
the companion star probably do not lead to significant transport in
the inner regions of the disks of close binaries because of their
relatively large Mach numbers. This would be particularly true of the
cold, quiescent disks considered here. However, no study to date has
considered a realistic, steeply increasing profile of surface density
with radius, as predicted for the quiescent disks by disk instability
models{\it(\Ref{can93})}. Because such a profile favors
tidally-induced transport in the inner regions of the
disk{\it(\Ref{sav94})}, this question remains open.

The concept of accretion driven by MHD turbulence during outburst and
by tidal perturbations during quiescence has several interesting
implications. It complicates the modeling of the disks of transient
close binaries, because, in quiescence, $\alpha$ is plausibly
non-uniform with radius, and becomes time-variable as the disk shrinks
and couples less and less with time to the tidal perturbation (a
minimum value for the disk outer radius is set by the specific angular
momentum of the accreted gas). It also allows for an efficiency of
transport in quiescent disks which differs from system to system
because it depends on the binary star parameters.\footnote{$^4$}{This
could account for a value of $\alpha_{\rm cold}$ in the quiescent disk
of the low $q$ dwarf nova WZ Sge which is orders of magnitude less
than what is usually inferred for other dwarf novae{\it(\Ref{sma93})}}
Finally, it implies that, in the absence of a massive companion,
weakly-ionized disks must rely on yet another transport mechanism or
they could be unable to accrete. Self-gravity is a plausible candidate
for angular momentum transport in the disks of T-Tauri stars and
active galactic nuclei, which are more massive than the disks
considered here. However, in systems with disks similar to those of
close binaries, such as supernova fallback disks or the disk which
lead to the planetary system around the pulsar PSR 1257+12, the
absence of a massive companion could result in essentially passive
disks.

\references {\parskip=0pt

\refis{pri81} J. Pringle, {\it Ann. Rev. Astron. Astrophys.} {\bf 19},
137 (1981).
\medskip \par

\refis{fkr92} J. Frank, A. R. King, D. J. Raine, {\it Accretion Power
in Astrophysics} (Cambrigde University Press, Cambridge, ed. 2, 1992).
\medskip \par

\refis{ss73} N. I. Shakura, R. A. Sunyaev, {\it Astron. Astrophys.}
{\bf 24}, 337 (1973).
\medskip \par

\refis{bh91} S. A. Balbus, J. F. Hawley, {\it Astrophys. J.} {\bf 376},
214 (1991).
\medskip \par

\refis{bh99} S. A. Balbus, J. F. Hawley, {\it Rev. Mod. Phys.}  {\bf
70}, 1 (1998).
\medskip \par

\refis{war95} B. Warner, {\it Cataclysmic variable stars}, (Cambridge
University Press, Cambridge, 1995).
\medskip \par

\refis{meys81} F. Meyer, E. Meyer-Hofmeister, {\it Astron. Astrophys.}
{\bf 104}, L10 (1981).
\medskip \par

\refis{sma84} J. Smak, {\it Acta Astron.} {\bf 34}, 161 (1984).
\medskip \par

\refis{can93} J. K. Cannizzo, in {\it Accretion Disks in Compact
Stellar Systems, Advanced Series in Astrophysics and Cosmology,
vol. 9}, J.C. Wheeler, Ed., p. 6 (1993).
\medskip \par

\refis{hgb96} J. F. Hawley, C. F. Gammie, S. A. Balbus, {\it
Astrophys. J.} {\bf 464}, 690 (1996).
\medskip \par

\refis{flem00} T. P. Fleming, J. M. Stone, J. F. Hawley, {\it
Astrophys. J.}, {\bf 530}, 464 (2000).
\medskip \par

\refis{gm98} C. F. Gammie, K. Menou, {\it Astrophys. J.} {\bf 492},
L75 (1998).
\medskip \par

\refis{ham98} J.-M. Hameury, K. Menou, G. Dubus, J.-P. Lasota,
 J.-M. Hur\'e, {\it Mon. Not. R. Astron. Soc.} {\bf 298}, 1048 (1998).
\medskip \par

\refis{xrb95} W. H. G. Lewin, J. van Paradijs, E. P. J. van den
Heuvel, {\it X-ray binaries} (Cambridge University Press, Cambridge,
1995).
\medskip \par

\refis{men00} K. Menou, J.-M. Hameury, J.-P. Lasota,
 R. Narayan, {\it Mon. Not. R. Astron. Soc.} {\bf 314}, 498 (2000).
\medskip \par

\refis{pp77} J. Papaloizou, J. E. Pringle, {\it
Mon. Not. R. Astron. Soc.} {\bf 181}, 441 (1977).
\medskip \par

\refis{gt80} P. Goldreich, S. Tremaine, {\it Astrophys. J.} {\bf 241},
425 (1980).
\medskip \par

\refis{ls91} M. Livio, H. Spruit, {\it Astron. Astrophys.}  {\bf 252},
189 (1991).
\medskip \par

\refis{rkcat99} H. Ritter, U. Kolb, {\it Astron. Astrophys. Suppl.} {\bf
129}, 83 (1998) [via the Vizier online catalogue].
\medskip \par

\refis{king88} A. R. King, {\it Quart. J. R. Astron. Soc.} {\bf 29}, 1
(1988).
\medskip \par

\refis{kkl96} A. R. King, U. Kolb, L. Burderi, {\it Astrophys. J.}
{\bf 464}, L127 (1996).
\medskip \par

\refis{sav94} G. J. Savonije, J. C. B. Papaloizou, D. N. C. Lin, {\it
Mon. Not. R. Astron. Soc.} {\bf 268}, 13 (1994).
\medskip \par

\refis{sma93} J. Smak, {\it Acta Astron.}  {\bf 43}, 101 (1993).
\medskip \par

\refis{ack}  ACKNOWLEDGEMENTS. The author is grateful to Steven
Balbus, Charles Gammie, Jeremy Goodman, Brad Hansen and Ramesh Narayan
for useful discussions. Support for this work was provided by NASA
through Chandra Postdoctoral Fellowship grant number PF9-10006 awarded
by the Chandra X-ray Center, which is operated by the Smithsonian
Astrophysical Observatory for NASA under contract NAS8-39073.

}
\endreferences
\endmode
\bigskip

\vfill
\eject

\psfig{figure=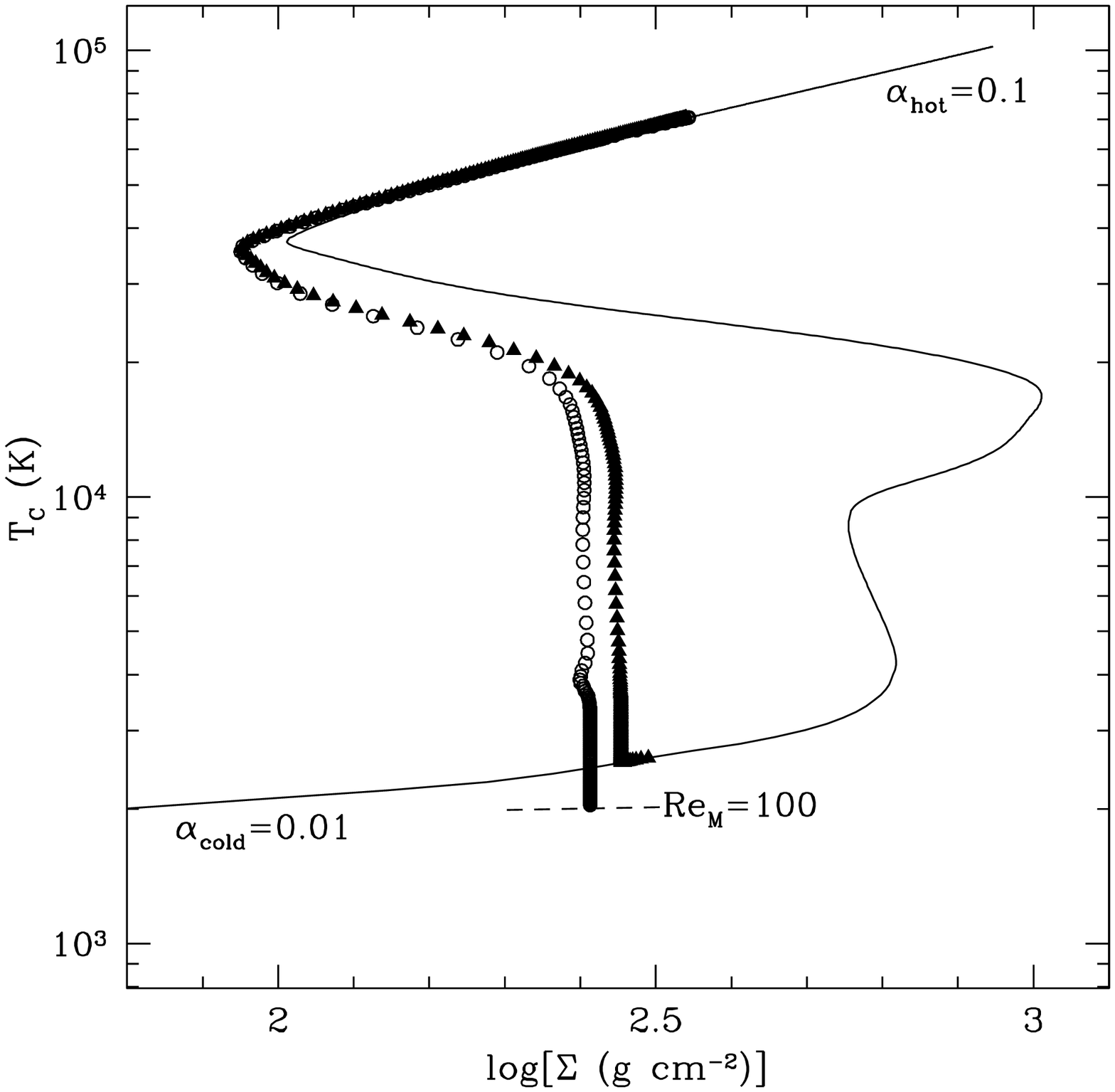,width=3.5in,height=3.5in}

\bigskip

\noindent{\ssbig{\bf Fig.~1.} The evolution into quiescence of a disk
annulus located at $2 \times 10^{10}$ cm from a central white dwarf is
shown in a surface density ($\Sigma$) vs. central temperature ($T_c$)
diagram. The solid line represents the disk thermal equilibria.  The
middle section, which corresponds to partially ionized gas, is
unstable and forces the annulus to a cyclic behavior.  The triangles
represent the evolution of the annulus, from top to bottom, in a
standard model with arbitrarily chosen values of the viscosity
parameter $\alpha$ on the stable branches ($\alpha_{\rm hot}=0.1$,
$\alpha_{\rm cold}=0.01$). The circles represent the evolution in a
model where the value of $\alpha$ is consistently taken as a function
of $Re_{\rm M}$ in the low ionization regime. In this case, the
annulus experiences a runaway cooling, down to values of $Re_{\rm M} <
100$. At such a low level of ionization, MHD turbulence dies away.}

\vfill
\eject

\psfig{figure=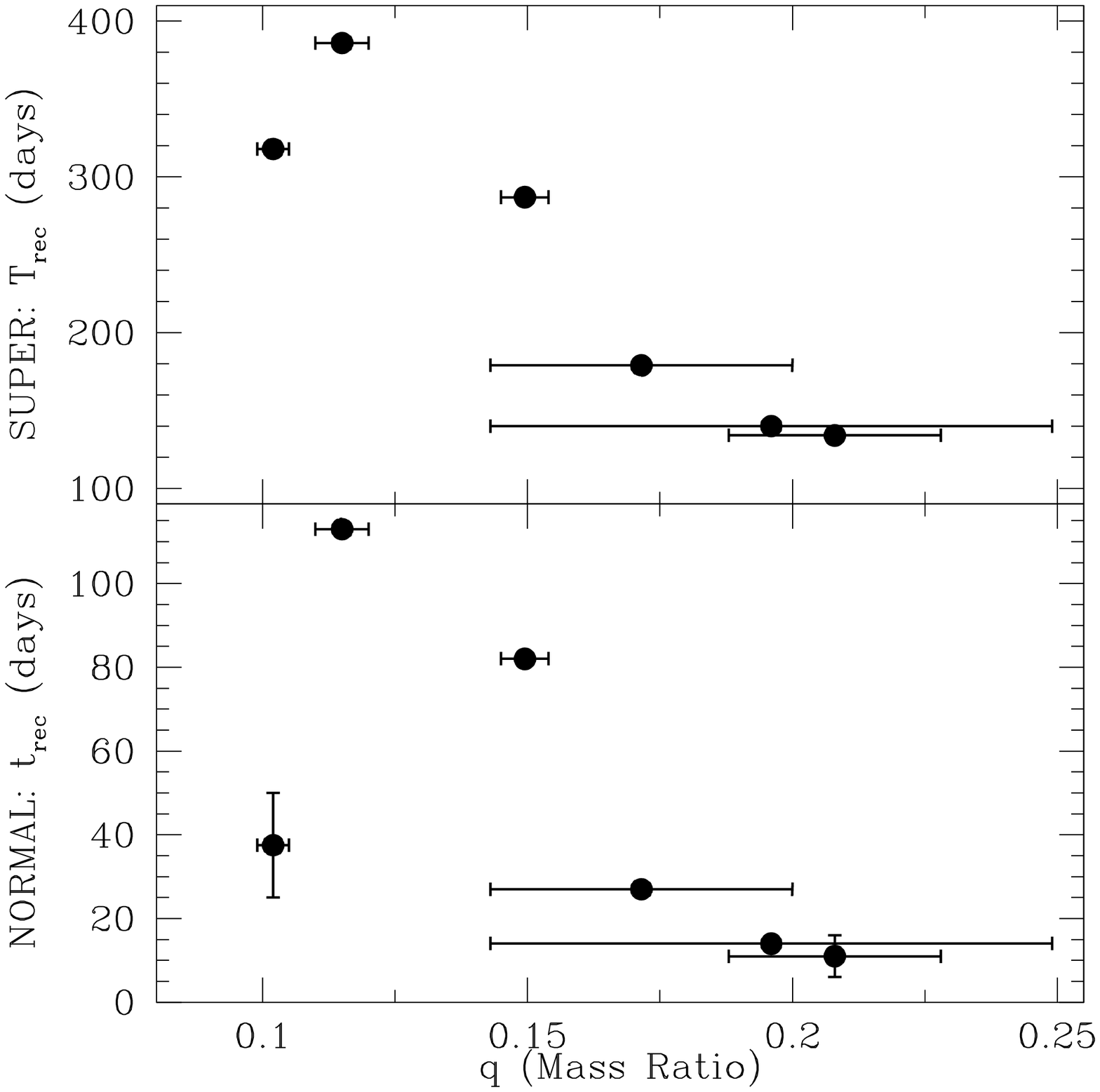,width=3.5truein,height=3.5truein}

\bigskip

\noindent{\ssbig{\bf Fig.~2.} The recurrence times of normal and super
outbursts of a sample of well-studied SU UMa stars are shown as a
function of their mass ratio. There is an indication for systems with
longer recurrence times having lower mass ratios. From left to right,
the six systems are OY Car, CU Vel, Z Cha, VW Hyi, WX Hyi and YZ Cnc,
with white dwarf masses $M_1/M_\odot =$0.68, 1.23, 0.84, 0.63, 0.9 and
0.82, respectively. The data are taken from the Ritter \& Kolb
catalogue{\it(\Ref{rkcat99}}). Masses are usually determined via
time-resolved photometric and spectroscopic methods.}

\vfill\eject

\bye